\documentclass[aps,prd,showpacs,onecolumn,nofootinbib,superscriptaddress,amsmath,amssymb]{revtex4}
\usepackage{epsfig}
\usepackage{graphicx}
\usepackage{color}
\usepackage{mathrsfs}
\usepackage{bm}
\newcommand{\rev}[1]{{#1}}

\usepackage{tikz}
\usetikzlibrary{decorations.pathmorphing}
\tikzset{zigzag/.style={decorate,decoration=zigzag}}

\begin{document}
\title{Feynman Rules in terms of the Wigner transformed Green functions}
\author{C. X. Zhang}\email{zhang12345s@sina.com}
\affiliation{Physics Department, Ariel University, Ariel 40700, Israel}

\author{M. A. Zubkov \footnote{On leave of absence from NRC "Kurchatov Institute" - ITEP, B. Cheremushkinskaya 25, Moscow, 117259, Russia}}
\email{zubkov@itep.ru}
\affiliation{Physics Department, Ariel University, Ariel 40700, Israel}

\begin{abstract}
In the models defined on the inhomogeneous background the propagators depend on the two space - time momenta rather than on one momentum as in the homogeneous systems. Therefore, the conventional Feynman diagrams contain extra integrations over momenta, which complicate calculations. We propose to express all amplitudes through the Wigner transformed propagators. This approach allows us to reduce the number of integrations. As a price for this the ordinary products of functions are replaced by the Moyal products.  The corresponding rules of the diagram technique are formulated using an example of the model with the fermions interacting via an exchange by scalar bosons. The extension of these rules to the other models is straightforward. This approach may simplify calculations in certain particular cases. The most evident one is the calculation of various non - dissipative currents.
\end{abstract}

\maketitle


\section{Introduction}

External fields provide the nontrivial inhomogeneous background to the field theoretical models that describe various physical systems both in the high energy physics
and in the condensed matter physics. Strong magnetic fields appear in the description of the early universe \cite{Vachaspati_1991}, of the neutron stars \cite{Reisenegger_2003} and can be created in heavy-ion collision experiments \cite{Kharzeev_2008}.
In condensed matter physics magnetic fields cause a lot of interesting phenomena. Possibly, the most popular one is the quantum Hall effect (QHE), which includes the mysterious fractional quantum Hall effect (FQHE).
In quantum Hall systems, the Hall resistivity $R_H$ as a function of magnetic field $B$ possesses the plateaus in the presence of interactions between the electrons, and impurities \cite{vonKlitzing_1980,Laughlin_1981}. Elastic deformations in materials is  another kind of external field experienced by the charged carriers, and may also change the behavior of the electric transport \cite{Guinea_2010,Arias_2015,Castro-Villarreal_2017,Amorim_2016}.

In the presence of external fields, the translational invariance is broken.
Therefore, the two-point Green functions $G(x_1,x_2)$ can not be
expressed in the form of the function of $(x_1-x_2)$, and after the Fourier transformation  the Green function depends on the initial and the final momenta, that are not equal in general case. An alternative to the Fourier transformation is
Wigner transformation \cite{Wigner_1932,Groenewold_1946,Moyal_1949}.
The Wigner-transformed Green function $G_W(R,p)$ has certain advantages compared to the ordinary Fourier transform  $\tilde{G}(p_1,p_2)$. Expressions in terms of $G_W(R,p)$ are more concise. We will see below, that the corresponding Feynmann diagram technique contains the same amount of integrations over momenta as in the homogeneous theory. The price for this is the appearance of the Moyal products instead of the ordinary multiplications. Sometimes the resulting expressions for the physical quantities are more useful than those that are obtained using the conventional Feynmann diagrams. An example is given by the Hall conductivity, which is expressed through the Wigner transformed Green functions using the Moyal products. The corresponding expression is the topological invariant in phase space, i.e. its value is not changed under the smooth modification of the system (see \cite{Zubkov+Wu_2019}). The similar (but simpler) constructions were used earlier to consider the intrinsic anomalous Hall effect and chiral magnetic effect \cite{Zubkov2016}. It has been shown that the corresponding currents are proportional to the topological invariants in momentum space. This method allows to reproduce the conventional expressions  for Hall conductivity \cite{TKNN1982}, and to prove the absence of the equilibrium chiral magnetic effect.
Recently, the electron-electron interactions have been taken into account,
and it has been shown to all orders in perturbation theory that the Hall conductivity (averaged over the system area) is proportional to the same topological invariant, as in the presence of interactions\cite{Zhang+Zubkov2019,Zhang_2019_JETPL}. This proof works equally well for the homogeneous systems with intrinsic anomalous Hall effect, and the non - homogeneous systems with quantum Hall effect in the presence of varying magnetic field, and varying electric potential of impurities.

In the theories with interactions, Feynman diagrams and Feynman rules are necessary to
evaluate various Green functions \cite{Feynman_1949, Feynman_1985}, and various physical amplitudes.
{ Each particular Feynman diagram ${\cal F}(X^{(f)} | X^{(i)})$ with $2N$ external fermion lines depends on the set of
input coordinates $X^{(i)}=\{x^{(i)}_a | a = 1,..., N\}$ and output coordinates $X^{(f)}=\{x^{(f)}_a | a = 1,..., N\}$.
The fermion lines pass through the whole diagram, and connect the  input/output coordinates into pairs
$X_a= (x^{(i)}_a, x^{(f)}_a)$ with $a = 1,..., N$.
If we denote the Wigner-transformed  function of  ${\cal F}(X^{(f)} | X^{(i)})$ by ${\cal F}(R | P)$,
where $R=\{r_a=(x^{(f)}_a + x^{(i)}_a)/2| a = 1,..., N\}$ and $P=\{p_a | a = 1,..., N\}$, while $p_a$ are the conjugate momenta with respect to $x^{(f)}_a-x^{(i)}_a$,
then the relation between  ${\cal F}(X^{(f)} | X^{(i)})$ and  ${\cal F}(R | P)$ is
$$
 {\cal F}(X^{(f)} | X^{(i)}) = \int \frac{d^{ND}P}{(2\pi)^D}
                                          e^{i \sum_a p_a (x^{(f)}_a-x^{(i)}_a)} {\cal F}_W(R| P),
$$
where $D$ is the dimension of space - time. The physical Green function ${\cal G}(x^{(f)}_a | x^{(i)}_a)$
corresponding to  $N$ incoming fermions, and $N$ outgoing fermions may be expressed through
the Feynman diagrams as the sum over permutations $\Pi$ of the sequence $(a=1,...,N)$:
$$
{\cal G}(X^{(f)} | X^{(i)}) = \sum_{\Pi} (-1)^{{\cal P}(\Pi)} {\cal F}(X^{(f)}_{\Pi} | X^{(i)})
= \sum_{\Pi}(-1)^{{\cal P}(\Pi)} \int \frac{d^{ND}P}{(2\pi)^{ND}}
e^{-i \sum_a p_a (x^{(f)}_{\Pi(a)}-x^{(i)}_a)} {\cal F}_W\Big( (X^{(f)}_{\Pi} + X^{(i)})/2| P \Big).
$$
}
Here $(-1)^{{\cal P}(\Pi)}$ is the parity of the permutation $\Pi$.
$X^{(f)}_{\Pi}$ is the set of permutated output coordinates $\{x^{(f)}_{\Pi(a)} | a = 1,..., N\}$
according to the permutation $\Pi$.
In the present paper we construct the diagram technique for the calculation of
the Wigner transformed Feynmann diagrams  ${\cal F}_W (R| P)$.
This technique expresses these quantities through the Wigner transformed bare propagators. For definiteness we consider the model of one Dirac fermion interacting minimally with the scalar field. The inhomogeneity is introduced to the system through the external Abelian gauge fields that interact with the scalar field and the fermion. Those fields may depend arbitrarily on coordinates. Correspondingly, each propagator depends on two momenta rather than one one momentum (as in the case of the homogeneous system). The generalization of our construction to the more general case is straightforward.

The paper is organized as follows:
In Sect. 2 the lagrangian of the model under consideration is introduced and certain basic expressions for the Wigner transformed Green functions are presented.
 In Sect. 3, the Feynman rules are obtained for the diagrams,
which contain zero or two external fermion lines.
 In Sect. 4, the Feynman rules are obtained for the diagrams,
which contain two external fermion lines and several internal fermion loops.
 In Sect. 5, we discuss the extension of our construction to the diagrams with more than two external fermion lines.
 In Sect. 6 we end with the conclusions.


\section{Wigner transform and the model under consideration}

Wigner transform is also known as Weyl transform \cite{Weyl_1927}.
Originally, it was closely intertwined with the development of phase-space formulation of quantum mechanics
(also called "deformation quantization") in 1930's  (see \cite{Vassilevich_2008,Vassilevich_2015} and references therein).
Compared with the operator formulation of quantum mechanics, the phase-space formulation
deals with the ordinary functions of coordinates and momenta: the so-called Wigner
function $W(q,p)$, which is considered as a quantum counterpart of classical  distribution
in phase space $(q,p)$ \cite{Balazs_1984}.
\rev{Wigner distribution $W(q,p)$ is not a positive real function. Therefore, strictly speaking, it cannot be considered as a phase space analogue of the probability distribution. For example, the superposition of two Gaussian wavepackets gives $W(q,p)$ with varying signs \cite{Zurek_1991}. Despite this drawback, a fluid analog of the quantum entropy flux in phase space can be formulated using Weyl-Wigner formalism, i.e. in terms of the Wigner function $W(q,p)$  \cite{Bernardini_2017}.
For positive defined Wigner distribution the analogue of von Neuman entropy may be defined as  \cite{Bernardini_2019,Wlodarz_2003}
$
S_{vN}=-\int dqdp W(q,p)\, {\rm ln}\, W(q,p).
$
Besides, Wigner distribution provides an expression for the quantum purity  \cite{Bernardini_2017}
$
{\cal P}=2\pi \int dqdp W(q,p)^2.
$
For pure states, ${\cal P}=1$, while for mixed states, ${\cal P}<1$.
Therefore $1-{\cal P}$ gives measure of the distance to the pure state.
}

In addition to its applications in quantum mechanics
\cite{Curtright_1998,Bastos_2008,Bernardini_2015,Bernardini_2017},
the Wigner transform has been extended to quantum field theory and has
been adopted to the investigation of the high energy physics and the many-body effects,
including various anomalous transport phenomena such as the quantum Hall effect,
the chiral separation effect, and the chiral vortical effect
\cite{Lorce_2011,Lorce_2012,Buot_1990,Buot+Jensen_1990,Miransky+Shovkovy_2015,Prokhorov+Teryaev_2018}.
In this section we will go along this line to introduce
 the Wigner transformation of the Green function.
Our starting point is the following lagrangian for the Dirac fermion interacting with the scalar field (in the presence of the inhomogeneous background given by the external hauge fields):
\begin {eqnarray}\label{Yukawa}
\mathcal{L}=\bar{\psi} (( i \partial_{\mu}-A_{\mu})\gamma^{\mu} - m) \psi
                  +(i \partial_{\mu} - B_{\mu})\phi (i \partial^{\mu} - B^{\mu})\phi - m^2_{\phi}\phi^2
                  -g \bar{\psi} \psi \phi
\end{eqnarray}
where $A_{\mu}$ and $B_{\mu}$ are the vector potentials of the external fields, which
can be different one from another.

For the fermions the two - point Green function $G$ satisfies equation $\hat{Q}(x_1) G(x_1,x_2)=\delta(x_1-x_2)$,
where $\hat{Q}(x)=( i \partial_{\mu}-A_{\mu}(x))\gamma^{\mu} - m $.
The Wigner transformation of $G$ is defined as
\begin {eqnarray}\label{Wigner}
G_W (R,p)=\int dr G(R+r/2, R-r/2) e^{-ipr}.
\end{eqnarray}
It satisfies the Groenewold equation $Q_W(R,p)\star G_W (R,p)=1$  \cite{Zubkov2016,Suleymanov_2019},
where $Q_W$ is the Weyl symbol of
operator $\hat{Q}$, while
$\star=e^{i(\overleftarrow\partial_R\overrightarrow\partial_p - \overleftarrow\partial_p\overrightarrow\partial_R) /2}$ is the Moyal product.

Similarly for the bosonic field $\phi$ operator
$\hat{U}(x)=( i\partial_{\mu}-B_{\mu}(x)) ( i\partial^{\mu}-B^{\mu}(x))- m^2_{\phi} $ is the inverse bare propagator,
and the Wigner - transformed Green function $D_W$ satisfies $U_W(R,p) \star D_W (R,p)=1$.
An important result of Wigner - Weyl calculus is  \cite{Suleymanov_2019,Bayen_1978,Littlejohn_1986}
\begin{eqnarray}\label{product}
C(x_1,x_2)=\int A(x_1,y)B(y,x_2)dy \Rightarrow
C_W(R,p)= A_W(R,p)\star B_W(R,p).
\end{eqnarray}
This result and its consequences will be used frequently in the further text.

\section{Feynman rules for the self energy and the fermion bubbles}
\label{sectbubble}
In this section we construct the Feynman rules for those diagrams corresponding to the fermion self energy \cite{Peskin}, in which there are no internal fermion loops.
We use the Wigner-transformed bosonic propagators $D^{(j)}$ and fermion propagators $G_{a}$. Here indices $j$ and $a$ enumerate the boson and the fermion lines correspondingly  entering the Feynman diagram.
Before the formulation of the Feynman rules, we introduce several auxiliary mathematical results.
\begin{enumerate}

\item{}
Our first auxiliary formula is as follows:
\begin{equation}
C(x_1,x_2)=\int A(x_1,y)H(y)B(y,x_2)dy \Rightarrow
C_W(R,p)= A(R,p)\star H(R)\star B(R,p)\label{f1}
\end{equation}
This result may be proven directly using Eq. (\ref{product}). Notice, that the Moyal product is associative, i.e.
$$
\Big(A(R,p) \star A(R,p)\Big) \star C(R,p) = A(R,p) \star \Big( A(R,p) \star C(R,p)\Big)
$$
This allows to omit the brackets in Eq. (\ref{f1}).

\item{}
The application of the Moyal product gives rise to the following expressions
\begin {eqnarray}\label{lemma1}
e^{ikR}\star G_{W}(R,p)= e^{ikR} G_{W}(R,p-k/2)     \\
G_{W}(R,p)\star e^{ikR} = e^{ikR} G_{W}(R,p+k/2)
\end{eqnarray}
and
\begin {eqnarray}\label{lemma2}
(A(R,p)e^{ikR})\star B(R,p)= [A(R,p)\star B(R,p-k/2)] e^{ikR}    \\
A(R,p)\star (e^{ikR} B(R,p))= [A(R,p+k/2)\star B(R,p)] e^{ikR}
\end{eqnarray}

\item{}
The above results give rise to the following
{\bf Lemma}
\begin {eqnarray}\label{Theorem}
&&G_1(R,p)\star e^{ik_1R}\star G_2(R,p)\star ...\star e^{ik_n R}\star G_{n+1}(R,p) \\
&&\overset{def}{=} G_1(R,p)\prod^{n}_{i=1} {}^{\star} (e^{ik_i R}\star G_{i+1}(R,p))  \\
&&= [\prod^{n}_{i=1} {}^{\star} G_{i}(R,p+p_i/2)] \quad e^{i\sum_j^n k_j R},
\end{eqnarray}
where $p_m=-\sum_{j=1}^{m-1} k_j +\sum_{j=m}^{n} k_j$.
\end{enumerate}

\begin{figure}[h]
	\centering  %
	\includegraphics[width=3cm]{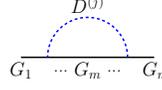} \vspace{0.5cm}
	\caption{The schematic representation of the diagrams of the fermionic self-energy without internal fermion loops. The solid line represents the fermion while
the dashed line represents the scalar. $G_i$ and $D^{(j)}$ are the fermionic and the bosonic Green functions
respectively. Dots stand for the additional ejections and absorbtions of the scalar by the fermion (those that are not shown explicitly).}  %
	\label{fig.1}   %
\end{figure}

In the Feynmann diagrams of the theory with the lagrangian of Eq.(\ref{Yukawa}) the two-point Green functions (propagators) are typically
represented by the solid lines (fermions), and the dashed lines (bosons).
After the Fourier transformation the bosonic propagator $\tilde D(k_{a},k_{b})$ in the presence of external field becomes the function of the two momenta, and the two exponential factors appear.
Let us take $\tilde D^{(j)}(k_{ja},k_{jb})$ in Fig. \ref{fig.1}
as an example. The propagator $D^{(j)}$ at its left end produces the factor $e^{ik_{ja}R}$,
and at the right end the factor $e^{ik_{jb}R}$, after the Fourier transform.

The positions of the fermionic propagator $G_m$ with respect to the given dash (the given bosonic propagator) are divided into three cases:
\begin{enumerate}

\item{} Both ends of the dashed line are right to $G_m$.

\item{} Both ends of the dashed line are left to $G_m$.

\item{} One end of the dashed line is on the left from $G_m$, and
the other end of the dashed line is right to $G_m$.

\end{enumerate}
Within the expression for the Wigner transformed self energy the influence of $D^{(j)}$ on various fermion propagators (better to say - on their Wigner transformations) $G_m$ ($1\leq m \leq n$) is as follows
\begin {eqnarray}\label{GreenStar1}
&&G_1(R,p+q_j/2)\star ... \star G_s(R,p+q_j/2)\star         \nonumber\\
&&G_{s+1}(R,p-k_j)\star ... \star G_{t-1}(R,p-k_j)\star    \nonumber\\
&&G_{t}(R,p-q_j/2)\star ... \star G_{n}(R,p-q_j/2)
\end{eqnarray}
where $q_j=k_{ja}-k_{jb}$, and $k_j=(k_{ja}+k_{jb})/2$. Here symbols of the Wigner transformation are omitted for brevity.
The corresponding expression for the Wigner transformation of the given self energy diagram becomes (we represent here only one bosonic propagator, and its influence on the Wigner transformed fermionic Green functions, the exponential factors coming from the other bosonic propagators are hidden inside the dots):
\begin {eqnarray}\label{GreenStar2}
\int [G_1(R,p) ... \star G_s(R,p)\circ_j\star
G_{s+1}(R,p-k_j)\star ... \star G_{t-1}(R,p-k_j)\star_j\circ
G_{t}(R,p)\star ...  G_{n}(R,p)] D^{(j)}(R,k_j)dk_1 ... dk_j ...
\end{eqnarray}
where $\circ_j=e^{-i\overleftarrow\partial_p\partial^{(j)}_R /2}$ and
${}_j\circ =e^{i \partial^{(j)}_R \overrightarrow\partial_p /2}$.
$\partial^{(j)}_R$ acts on $D^{(j)}$ only. {The right derivative  $\overrightarrow\partial_p$ acts on all propagators standing right to the symbol ${}_j\circ$. The left derivative  $\overleftarrow\partial_p$ acts on all propagators standing left to the symbol $\circ_j$.  }

\begin{figure}[h]
	\centering  %
	\includegraphics[width=7cm]{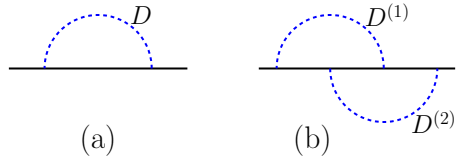}\vspace{1cm}  %
	\caption{(a)One-loop Feynmann diagram for the self energy.
     (b)An entangled two-loop Feynmann diagram for the self energy.}  %
	\label{fig.2}   %
\end{figure}

\rev{In order to better understand the computation rules,
let us consider the specific examples shown in Fig.\ref{fig.2}.
Fig.\ref{fig.2}(a) shows the leading order contribution to the
Green function from the Yukawa interaction, which is the simplest case of
the interacting fermionic Green function.
The corresponding expression is given by
\begin {eqnarray}\label{Green_1st}
\int [G_1(R,p) \circ_D \star G_2(R,p-k)\circ_D \star G_1(R,p)] D_W(R,k) dk
\end{eqnarray}
Fig.\ref{fig.2}(b) is a more complicated case, in which the two loops
entangle with each other.}
The solid line is separated into 5 segments by the two dashes.
The second and the third segments are in "parallel" with $D^{(1)}$,
therefore, their momentum variables include $k_1$. Operators $\circ_1$
and ${}_j \circ$ are inserted before and after these segments.
Similarly, the propagator $D^{(2)}$ affects the third and the fourth
segments.
Finally, the corresponding expression of the Feynmann diagram of Fig.\ref{fig.2}(b) is given by
\begin {eqnarray}\label{GreenStar_eg}
\int\int && [G_1(R,p) \circ_1 \star G_2(R,p-k_1)\circ_2 \star  G_3(R,p-k_1-k_2)\star {}_{1}\circ
G_4(R,p-k_2)\star {}_{2}\circ G_5(R,p)]           \nonumber\\
&&  D^{(1)}_W(R,k_1) D^{(2)}_W(R,k_2) dk_1 dk_2.
\end{eqnarray}

\begin{figure}[h]
	\centering  %
	\includegraphics[width=5cm]{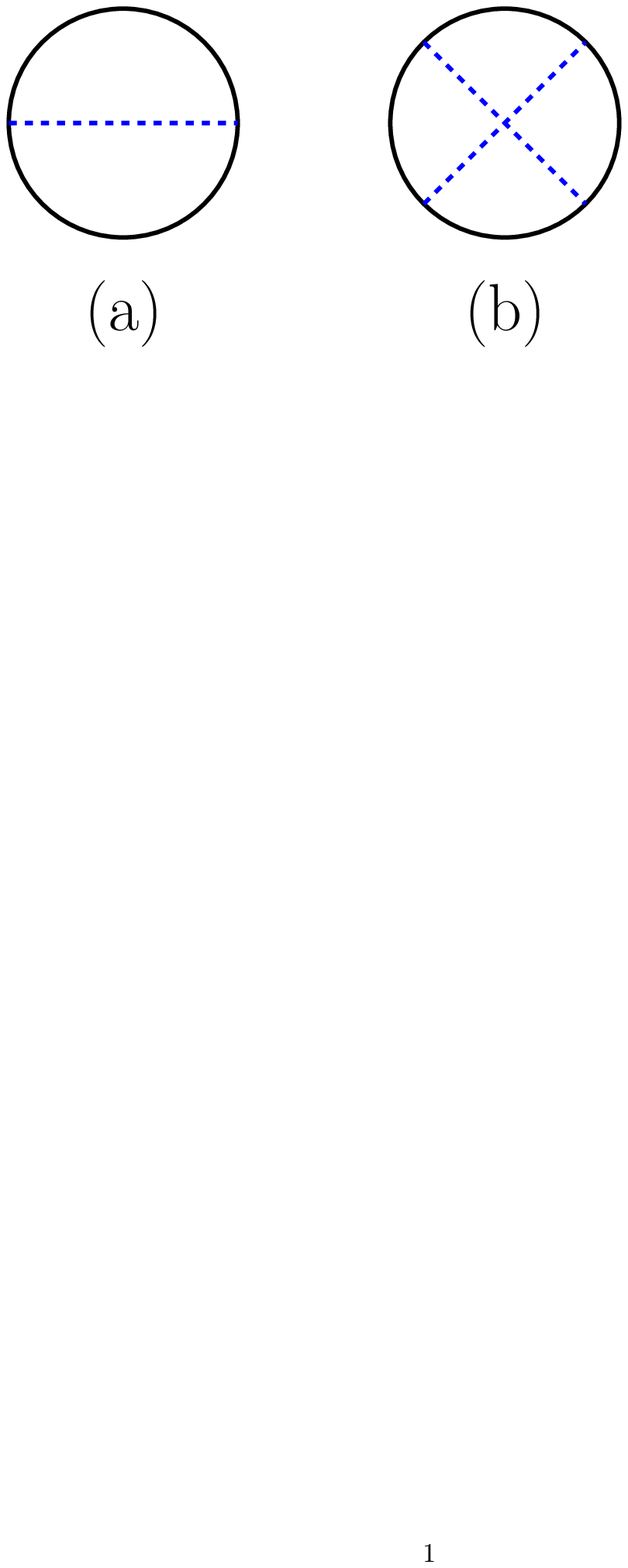}  
	\caption{Fermionic bubles.}  %
	\label{fig.3}   %
\end{figure}

The fermionic bubbles like those presented in Figs. \ref{fig.3} do not enter the expressions for the physical scattering amplitudes. However, they enter expressions for the thermodynamical potentials, and, moreover, are used for the proof that the Hall conductivity is not affected by weak interactions (see, e.g. \cite{Zhang_2019_JETPL}).

Below we represent the Feynman diagrams given on Fig. \ref{fig.3} in terms of the Wigner transformed propagators. The bubble (a) corresponds to expression
\begin {eqnarray}\label{bubble_a}
\frac{1}{2}\int Tr [G_W(R,p-k) \star {}_{1}\circ G_W(R,p)] D^{(1)}_W(R,k) dk.
\end{eqnarray}
Because of the trace, it also can be rewritten as
\begin {eqnarray}\label{bubble_a'}
\frac{1}{2}\int Tr [G_W(R,p) \circ_{1}\star G_W(R,p-k)] D^{(1)}_W(R,k) dk,
\end{eqnarray}
equivalently. As for bubble (b) the corresponding formula
is
\begin {eqnarray}\label{bubble_b}
\frac{1}{4}\int Tr [G_W(R,p-k_1) \circ_2 \star  G_W(R,p-k_1-k_2)
        \star {}_1\circ G_W(R,p-k_2) \star {}_2\circ G_W(R,p)]
D^{(1)}_{W}(R,k_1) D^{(2)}_{W}(R,k_2) dk_1 dk_2.
\end{eqnarray}
\rev{It is interesting to notice relation between the Feynman diagrams in
Fig.\ref{fig.2} and Fig.\ref{fig.3}. If one glues the two end points of
each diagram in Fig.\ref{fig.2}, one obtains the corresponding
diagram in Fig.\ref{fig.3}.
Such an observation will be useful in the applications of the proposed technique
discussed in Sec.\ref{sec_current}.}

%

We conclude this section with the formulation of the rules for the calculation of the  Feynman diagrams with two external fermion lines (the self energy), and without external fermion lines (the fermion bubbles). In both cases there should be no extra fermion loops.
\begin{enumerate}

\item{}

Label momenta $p$, $p-k_j$ ... within the graph, according to the "law of momentum conservation". This "law" means that we write down the momenta that would take place in the same diagram of the homogeneous theory. The combinatorial symmetry factor is to be added to each diagram. This factor is also identical to the one of the same diagram of the homogeneous theory.

\item{}

 Write down the series $G_W(R,p)\star G_W(R,p-k_j)\star ...$ along the
fermion line,  according to the labelled momenta of the graph. The inhomogeneity of the theory is now encoded in the dependence on $R$ while momenta entering these expressions are conserved as if we would deal with the homogeneous theory.

\item{}

Insert $\circ_j$ and ${}_j\circ$ to the series at the starting and ending
points of $D^{(j)}(R,k_j)$, according to the Feynman diagram. For the case of
the bubble, the trace is introduced, and the first $\circ_{D}\star$
operator is omitted.

\end{enumerate}

\section{The cases when the internal fermion loop is present}

\begin{figure}[h]
	\centering  %
	\includegraphics[width=5cm]{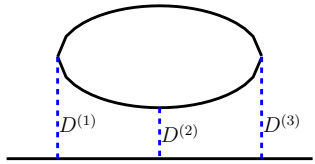} 
	\caption{An example of the  Feynmann diagram in self-energy, which contains two fermion lines. One of the fermion lines forms  an internal loop. }  %
	\label{fig.4}   %
\end{figure}

In the previous section, the Feynman  rules have been obtained
for the diagrams corresponding to the two-point Green functions (and the fermion bubbles), in which  only one fermion line is present. (This line passes through the whole diagram in the case of the fermion self - energy, and is closed to form the loop in the case of the fermion bubble.) In the present section, we consider the diagrams for the self - energy/fermion bubbles that include additional fermion loops. An example of such a diagram is presented in Fig. \ref{fig.4}.
\rev{We choose this diagram, because it is relatively general, and cannot be factorized.
Considering such an example we can figure out the general Feynman rules. }
The Feynman diagram in Fig. \ref{fig.4} may be evaluated as follows:
\begin {eqnarray}\label{Green_2line(1)}
{\cal F}(x_1|x_2) &=& \int G(x_1,y_1)G(y_1,y_2)G(y_2,y_3)G(y_3,x_2)
Tr[G(y_4,y_5)G(y_5,y_6)G(y_6,y_4)]\nonumber\\&&
D(y_1,y_4)D(y_2,y_5)D(y_3,y_6) dy_1 ... dy_6
\end{eqnarray}
Using relation $D(x,y)=\int e^{ik_a x}\tilde{D}(k_a,k_b)e^{-ik_b y} dk$,
and applying Wigner transformation, we obtain
\begin {eqnarray}\label{Green_2line(2)}
{\cal F}_W(R_1|p_1) &=&\int G_W(R_1,p_1+\frac{k_{1a}}{2}+\frac{k_{2a}}{2}+\frac{k_{3a}}{2})  \star
G_W(R_1,p_1-\frac{k_{1a}}{2}+\frac{k_{2a}}{2}+\frac{k_{3a}}{2}) \star  \nonumber\\
&&G_W(R_1,p_1-\frac{k_{1a}}{2}-\frac{k_{2a}}{2}+\frac{k_{3a}}{2}) \star
G_W(R_1,p_1-\frac{k_{1a}}{2}-\frac{k_{2a}}{2}-\frac{k_{3a}}{2})         \nonumber\\
&&Tr[G_W(R_2,p_2-\frac{k_{1b}}{2}-\frac{k_{2b}}{2}-\frac{k_{3b}}{2})  \star
G_W(R_2,p_2+\frac{k_{1b}}{2}+\frac{k_{2b}}{2}-\frac{k_{3b}}{2}) \star
G_W(R_2,p_2+\frac{k_{1b}}{2}+\frac{k_{2b}}{2}+\frac{k_{3b}}{2})
]                                                                   \nonumber\\
&&e^{ik_{1a} R_1}\tilde{D}(k_{1a},k_{1b})e^{-ik_{1b} R_2}
e^{ik_{2a} R_1}\tilde{D}(k_{2a},k_{2b})e^{-ik_{2b} R_2}
e^{ik_{3a} R_1}\tilde{D}(k_{3a},k_{3b})e^{-ik_{3b} R_2}\nonumber\\&&
 dk_{1a} dk_{2a} dk_{3a} dk_{1b} dk_{2b} dk_{3b} dR_2 dp_2
\end{eqnarray}
\rev{Notice that in the given integrals there are two groups of variables: $(R_1,p_1)$ and $(R_2,p_2)$, which correspond to the fermion line and the fermion loop, correspondingly.
This expression can be simplified introducing the Moyal product $\circ$ between the
fermionic Green functions and the bosonic ones, which leads to the following expression}
\begin {eqnarray}\label{Green_2line(3)}
{\cal F}_W(R_1|p_1) &=& G_W(R_1,p_1) \star \overset{(1,1)}{\circ} G_W(R_1,p_1) \star \overset{(2,1)}{\circ} G_W(R_1,p_1) \star \overset{(3,1)}{\circ} G_W(R_1,p_1)  \nonumber\\
&& \int Tr [ \overset{(1,2)}{\circ} G_W(R_2,p_2) \star \overset{(2,2)}{\circ} G_W(R_2,p_2) \star \overset{(3,2)}{\circ} G_W(R_2,p_2)]  \nonumber\\
&& D^{(1)}(R_1,R_2)D^{(2)}(R_1,R_2)D^{(3)}(R_1,R_2) dR_2 dp_2
\end{eqnarray}
where $\overset{(i,j)}{\circ}=exp(\frac{i}{2}({\partial}^{(i)}_{R_j} \overrightarrow{\partial}_{p_j}-\overleftarrow{\partial}_{p_j}{\partial}^{(i)}_{R_j}))$,
in which ${\partial}^{(i)}_{R_j} $ acts on $D^{(i)}(R_1,R_2)$ only. {The right derivative  $\overrightarrow\partial_{p_j}$ acts on all fermion propagators standing right to the symbol $\overset{(i,j)}{\circ}$. The left derivative  $\overleftarrow\partial_{p_j}$ acts on all propagators standing left to the symbol $\overset{(i,j)}{\circ}$.  }

This example may be easily extended to the diagram of general type with two or zero external fermion lines, and any number of internal fermion loops.

\section{Diagrams with more than two legs}

\begin{figure}[h]
	\centering  %
	\includegraphics[width=7cm]{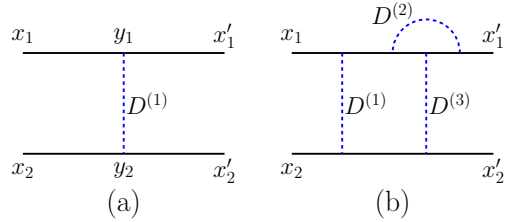} 
	\caption{(a) The simplest diagram with  four external fermion lines.
            (b) A more complicated example of the diagram with four external fermion lines.}  %
	\label{fig.5}   %
\end{figure}

Up to now, we only considered the two-point fermionic Green functions, which correspond to the Feynman diagrams with two legs formed by one fermion line, and the fermion bubbles without external legs. Our consideration may easily be generalized to the case of an arbitrary number of external fermion lines.
As an illustration let us consider the simple example shown in Fig.\ref{fig.5}(a).
In coordinate space the corresponding Feynman diagram is
\begin {eqnarray}\label{Green_4legs_a}
{\cal F}(x_1,x'_1|x_2,x'_2)=
\int G(x_1,y_1)G(y_1,x'_1)D(y_1,y_2)
     G(x_2,y_2)G(y_2,x'_2) dy_1 dy_2.
\end{eqnarray}
It has been mentioned in the Introduction, that we define the Wigner transformation of such a diagram that corresponds to the pairs $(x_1,x'_1)\rightarrow (R_1,p_1)$,
$(x_2,x'_2)\rightarrow (R_2,p_2)$. After some tedious algebra, one can get the final result for
${\cal F}_W(R_1,R_2|p_1,p_2)$. Extending the diagram technique of the previous section to this case we are able to write the corresponding formula directly:
\begin {eqnarray}\label{Green_4legs_b}
{\cal F}_W(R_1,R_2|p_1,p_2)=
 [G_W(R_1,p_1)\overset{(1,1)}{\circ} \star G_W(R_1,p_1)]
 [G_W(R_2,p_2)\overset{(1,2)}{\circ} \star G_W(R_2,p_2)] D^{(1)}(R_1,R_2)
\end{eqnarray}

Another example, which is more complicated, is shown in Fig.\ref{fig.5}(b). The corresponding
Wigner-transformed expression is
\begin {eqnarray}\label{Green_4legs_c}
{\cal F}_W(R_1,R_2|p_1,p_2)&=&\int dk
 [G_W(R_1,p_1)\overset{(1,1)}{\circ} \star G_W(R_1,p_1)
  \circ_2 \star G_W(R_1,p_1-k) \overset{(3,1)}{\circ}\star G_W(R_1,p_1-k) \star{}_2\circ G_W(R_1,p_1)]  \nonumber\\
&& [G_W(R_2,p_2)\overset{(1,2)}{\circ} \star G_W(R_2,p_2)\overset{(3,2)}{\circ} \star G_W(R_2,p_2)]
D^{(1)}(R_1,R_2)D_W^{(2)}(R_1,k)D^{(3)}(R_1,R_2)
\end{eqnarray}

\rev{Now let us summarize the general rules of the diagram technique illustrated by the above considered particular cases.
\begin{enumerate}
\item{ Fermi skeleton:}
for each fermion line $L_i$ (either closed or open), one needs a spatial coordinate $R_i$, and momenta $p_a$. Write down the series $G\star G...$ according to the rules presented at the end of
section \ref{sectbubble} (there instead of $R$ we insert $R_i$).
\item{Dashed lines connect points that belong to the same fermion line.}
The dashed lines, that start and end at the same fermion line result in the same operators $\circ_j$ and ${}_j\circ$ as in the previous section.
The dashed lines connecting different fermion lines are omitted in this step.
\item{Dashed lines connecting distinct fermion lines:}
for the boson propagators, whose ends belong to different fermion lines (denote those fermion line  $L_i$ and $L_j$) we use the boson propagator $D(R_i,R_j)$ in coordinate space rather than the Wigner transformed propagator $D_W$. Then the circle operators $\overset{(i,j)}{\circ}$ are inserted to the series $G\star G...$ at the positions of the ejection/absorbtion of the dashed line connecting $L_i$ and $L_j$. If necessary,
$D(R_i,R_j)$ may be expressed through $D_W$.
\item{} Combinatorial symmetry factors for each diagram are identical to those of the corresponding homogeneous theory.
\end{enumerate}}


\section{Application of the proposed technique}\label{sec_current}

\subsection{Bloch theorem}

\begin{figure}[h]
	\centering  %
	\includegraphics[width=7cm]{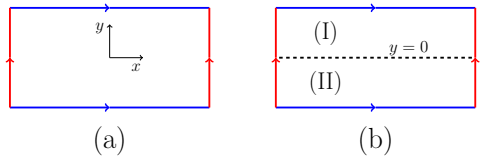}
	\caption{The two-dimensional fermion system located on the torus.
   The edges with the same color (and with the same direction of arrows) are glued
   together.}  %
	\label{fig_torus}   %
\end{figure}

In this section, we closely follow \cite{Zhang+Zubkov2019_Bloch}, where the particular case of the above given general diagram technique has been considered. In this particular case the above given Feynman rules are applied for the calculation of loop corrections to the persistent electric current in the system with periodic boundary conditions in space. This is the system defined on torus of infinite size (the two dimensional torus is represented in Fig.\ref{fig_torus}). We will discuss the model with lagrangian of Eq.(\ref{Yukawa}), in which the field $B$  is switched off.

\begin{figure}[h]
	\centering  %
	\includegraphics[width=5.5cm]{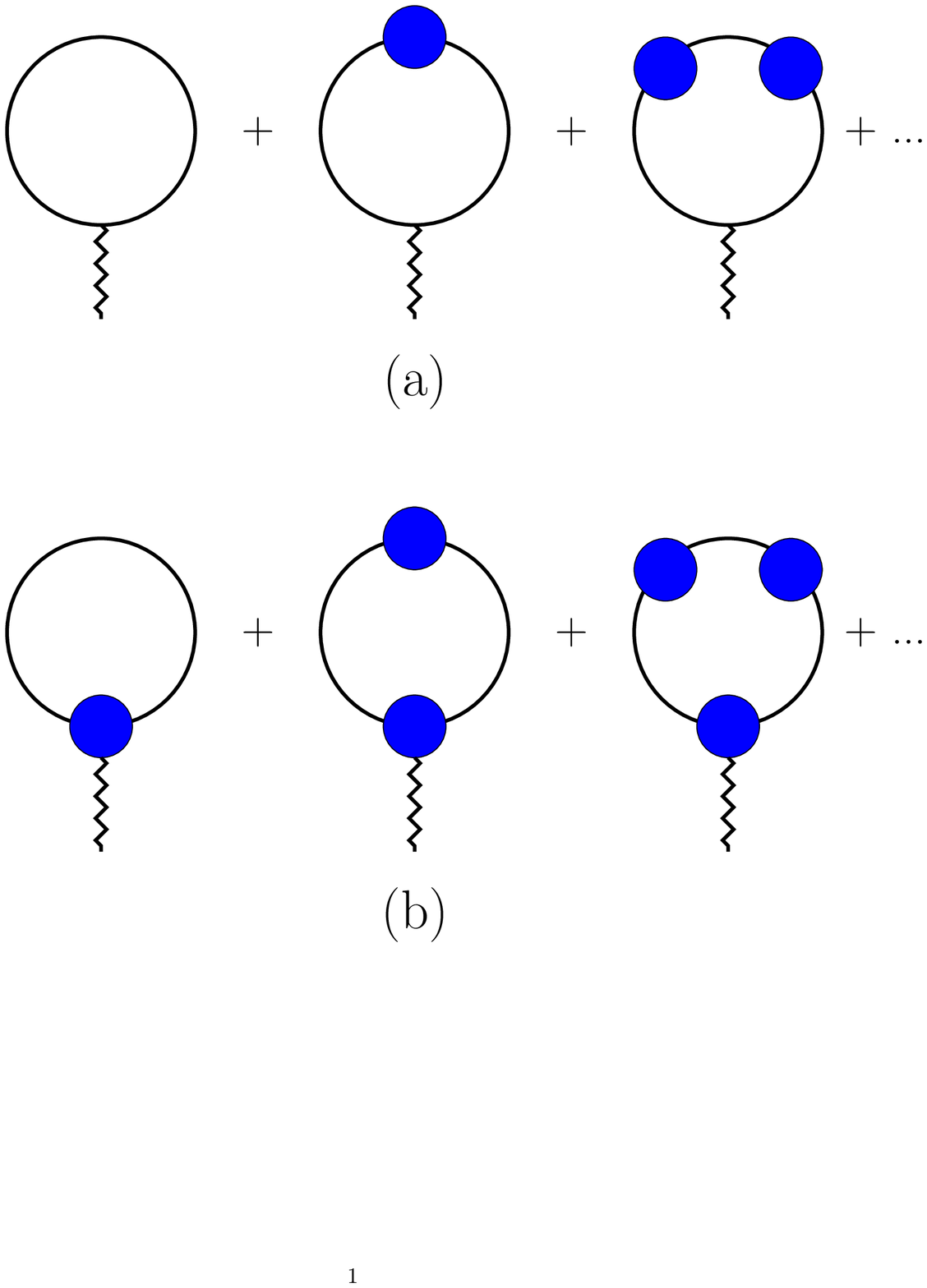} %
	\caption{(a) Feynmann diagrams  for the current $I(g) = \int_R \int_p Tr G_{g,W} \partial_{p_x} Q_{W}$.
       The filled circles mark $\Sigma_W$. The external wavy line marks the position of $\partial_{p_x} Q_{W}$.
         (b) Feynmann diagrams for $\Delta I(g) =  \int_R \int_p Tr G_{g,W} \partial_{p_x} \Sigma_{W}$.
        The filled circle with the external wavy line marks $\partial_{p_x} \Sigma_{W}$.}
     \label{Fig_tadpole}
\end{figure}

The bosonic two - point Green function is the function of the distance between the points $D(x_1-x_2)$.
Without loss of generality, we consider only the current along the $x$-axis.
The averaged current (as a function of the Yukawa coupling constant $g$)
is given by
\begin{eqnarray}\label{current}
I(g) = \int \frac{ d^3 R}{\beta S} \int \frac{d^3 p}{(2\pi)^3}
            {\rm Tr}\ G_{g,W}(R,p) \star  \frac{\partial}{\partial p_x} Q_{W}(R,p)
\end{eqnarray}
where integrations are over imaginary time, spatial coordinates,
imaginary frequency, and spatial momenta.
$\beta=1/T$, where $T$ is temperature (which is assumed to be small), $S$ is area of the system.
Here, $G_{g,W}=G_{W}+G_{W}\star  \Sigma_W \star G_{W}+...$
is the interacting Green function, $G_{W}$ is the non-interacting Green function,
and $\Sigma_W$ is the self-energy.
After inserting the expression of $G_{g,W}$ into Eq.(\ref{current}),
the current can be expressed as a series $I(g) =\sum_{n=0}^{\infty} I^{(n)}$ with
\begin {eqnarray}\label{current_component_a}
I^{(n)}=\int \frac{ d^3 R}{\beta S} \int \frac{d^3 p}{(2\pi)^3}  {\rm Tr}\ \big(G_{W} \star \Sigma_W \star\big)^n G_{W} \star  \frac{\partial Q_{W}}{\partial p_x}
\end{eqnarray}
The corresponding Feynman diagrams are shown in Fig.\ref{Fig_tadpole} (a)
It is easy to see that when $g=0$, $G_{g,W}$ is reduced to $G_W$, i.e. $G_{0,W}=G_W$,
which satisfies $G_W \star Q_W =1$.
Similarly, let us define $Q_{g,W}=Q_W - \Sigma_W$, then $Q_{0,W}=Q_W$ and
one can find that $G_{g,W}\star Q_{g,W} =1$.

Let us compare the expression for the averaged current
with the following expression written through
the interacting Green function $G_{g,W}$ and $Q_{g,W}$
\begin{equation}\label{tildeI}
\tilde{I}(g) =\int \frac{ d^3 R}{\beta S} \int \frac{d^3 p}{(2\pi)^3} Tr G_{g,W}(R,p) \star  \frac{\partial}{\partial p_x} Q_{g,W}(R,p).
\end{equation}
For this purpose we calculate the difference
$\Delta I(g) = I(g)-{\tilde I}(g)$,
which is given by
\begin {eqnarray}  \label{current_delta}
\Delta I =\int \frac{ d^3 R}{\beta S} \int \frac{d^3 p}{(2\pi)^3}   Tr G_{g,W}(R,p) \star  \frac{\partial}{\partial p_x} \Sigma_{W}(R,p)
\end{eqnarray}
and has the expansion $\Delta I= \sum_{n=0}^{\infty} \Delta I^{(n)}$, with
\begin {eqnarray}\label{current_component_b}
\Delta I^{(n)}=\int \frac{ d^3 R}{\beta S} \int \frac{d^3 p}{(2\pi)^3} {\rm Tr}\ \big(G_{W} \star \Sigma_W \star\big)^n G_{W} \star  \frac{\partial \Sigma_{W}}{\partial p_x}
\end{eqnarray}
The Feynmann diagrams corresponding to $\Delta I$ are represented in Fig. \ref{Fig_tadpole} (b).
It has been proven that $\Delta I^{(n)} = I^{(n+1)}$ \cite{Zhang+Zubkov2019_Bloch}, and therefore
\begin{equation}\label{deltaI}
\Delta I(g) = I(g)-I^{(0)} = I(g)-I(0)
\end{equation}
We find that the averaged current is given by the integral in Eq. (\ref{tildeI}),
as long as the value of the current remains equal to its value
without interactions, i.e. if $I(g)=I(0)$, then $ I(g)= {\tilde I}(g)$.

\begin{figure}[h]
	\centering  %
	\includegraphics[width=4cm]{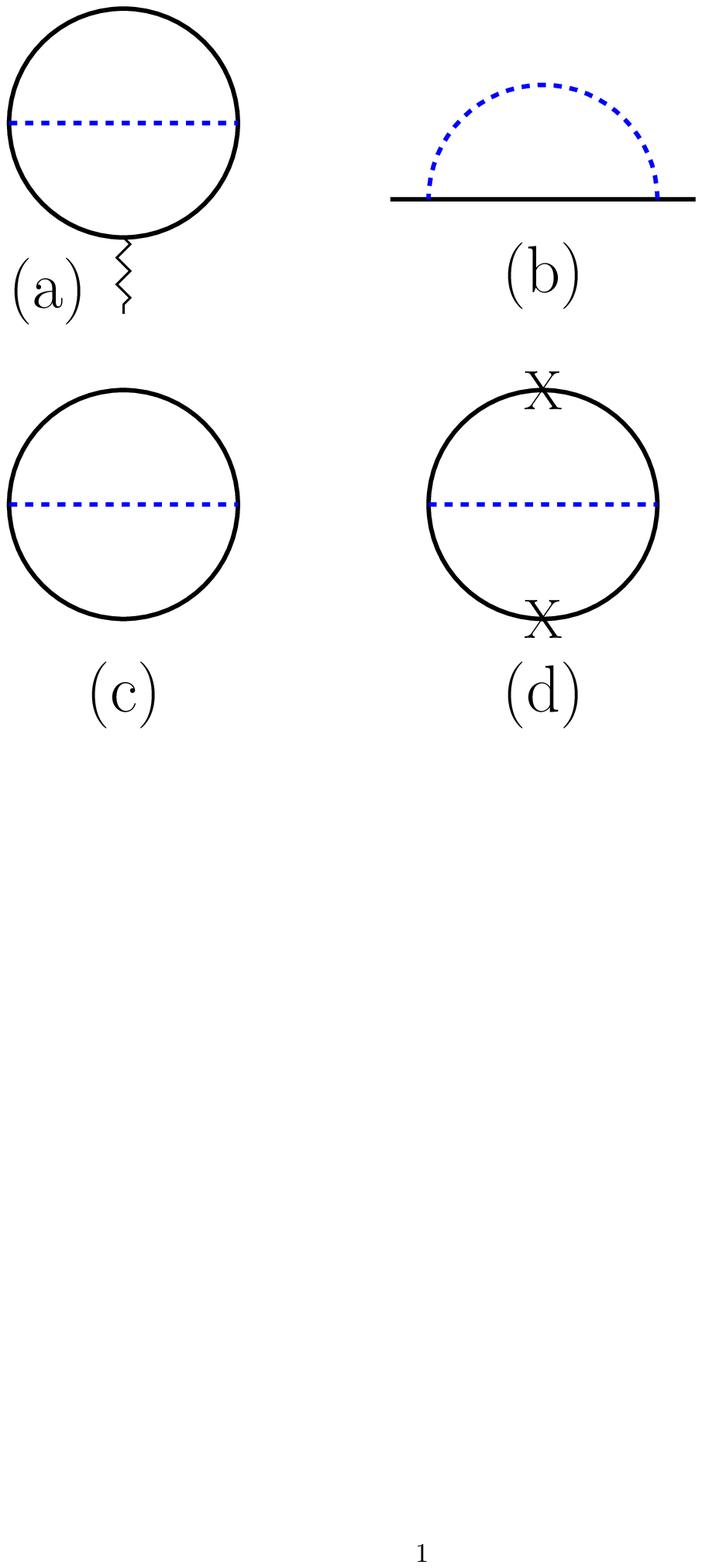}
	\caption{Feynman diagrams for the first order corrections to electric current:
(a) - the first order contribution to the current itself, (b) - the corresponding contribution to the self energy,
(c) is the "progenitor" diagram. Each cross in (d) being substituted by the derivative $\partial_p Q_{0,W}$ leads to the diagram of (a).
 }  %
	\label{fig_bubble_leading}   %
\end{figure}

It can be proved that indeed $I(g) = I(0)$ in the region of analyticity in $g$,
i.e. as long as the perturbation theory in $g$ may be used \cite{Zhang+Zubkov2019_Bloch}.
Here we present the part of the proof in the leading order $\sim O(g^2)$, i.e. we will show that
${\cal I}_1=0$.
The  leading order contribution from the Yukawa interaction is
depicted by the Feynman diagram shown in Fig.\ref{fig_bubble_leading}(a) and
the corresponding expression is
\begin {eqnarray}\label{current_1st}
{\cal I}_1 = -\int \frac{ d^3 R}{\beta S} \int \frac{d^3 p}{(2\pi)^3}  {\rm Tr}  \Big[ \int \frac{d^3 q}{(2\pi)^3} G_{W}(R,p-q){\cal D}(q) \Big]
                     \star  \frac{\partial}{\partial p_x} G_{W}(R,p)
\end{eqnarray}
Expression in the square brackets is the self-energy $\Sigma_W$ in the leading order,
which is shown in Fig.\ref{fig_bubble_leading}(b).
${\cal D}(q)$ is the bosonic Green function in momentum space, i.e. the Fourier transformation of $D(x_1-x_2)$.
In order to prove that ${\cal I}_1=0$, we consider the bubble diagram shown
in Fig.\ref{fig_bubble_leading}(c), which can be expressed as
\begin {eqnarray}\label{bubble_1st}
{\cal B}_1 = -\int \frac{ d^3 R}{\beta S} \int \frac{d^3 p}{(2\pi)^3}  \int \frac{d^3 q}{(2\pi)^3}
 {\rm Tr}   G_{W}(R,p-q) \star   G_{W}(R,p) {\cal D}(q).
\end{eqnarray}
Because of the integration over $p$, the operator $\partial_{p_x}$
being inserted into the integrand gives zero:
\begin {eqnarray}\label{bubble_1st_a}
-\int \frac{ d^3 R}{\beta S} \int \frac{d^3 p}{(2\pi)^3}  \int \frac{d^3 q}{(2\pi)^3}
 {\rm Tr} \frac{\partial}{\partial p_x}  \Big[  G_{W}(R,p-q) \star   G_{W}(R,p) \Big] {\cal D}(q)=0.
\end{eqnarray}
The operation $\partial_{p_x}$ produces two terms, marked by "crosses" in
Fig.\ref{fig_bubble_leading}(d).
After integration by parts and changing variables, we found that
each term gives the same result ${\cal I}_1$, as long as ${\cal D}(q)$ is an even function, i.e. ${\cal D}(-q)={\cal D}(q)$.
Therefore, $2 {\cal I}_1 =0$, and then ${\cal I}_1=0$, which shows that
the  leading order correction to the current is zero.

Notice, that the bubble diagram in Fig.\ref{fig_bubble_leading}(c) "generates" the Feynman diagram(s) under
discussion through the operation $\partial_{p_x}$:
after cutting Fig.\ref{fig_bubble_leading}(c) at the positions marked by crosses "X"
in Fig.\ref{fig_bubble_leading}(d) the
self-energy diagrams of Fig.\ref{fig_bubble_leading}(b) emerge,
which contribute to the current shown in Fig.\ref{Fig_tadpole}(a) and
Fig.\ref{fig_bubble_leading}(a).
Such a diagram  of Fig.\ref{fig_bubble_leading}(c) was called a "progenitor" in \cite{parity_anomaly}.
It generates the Feynman diagrams giving corrections to the current,
and the latter diagrams cancel each other.

\begin{figure}[h]
	\centering  %
	\includegraphics[width=5cm]{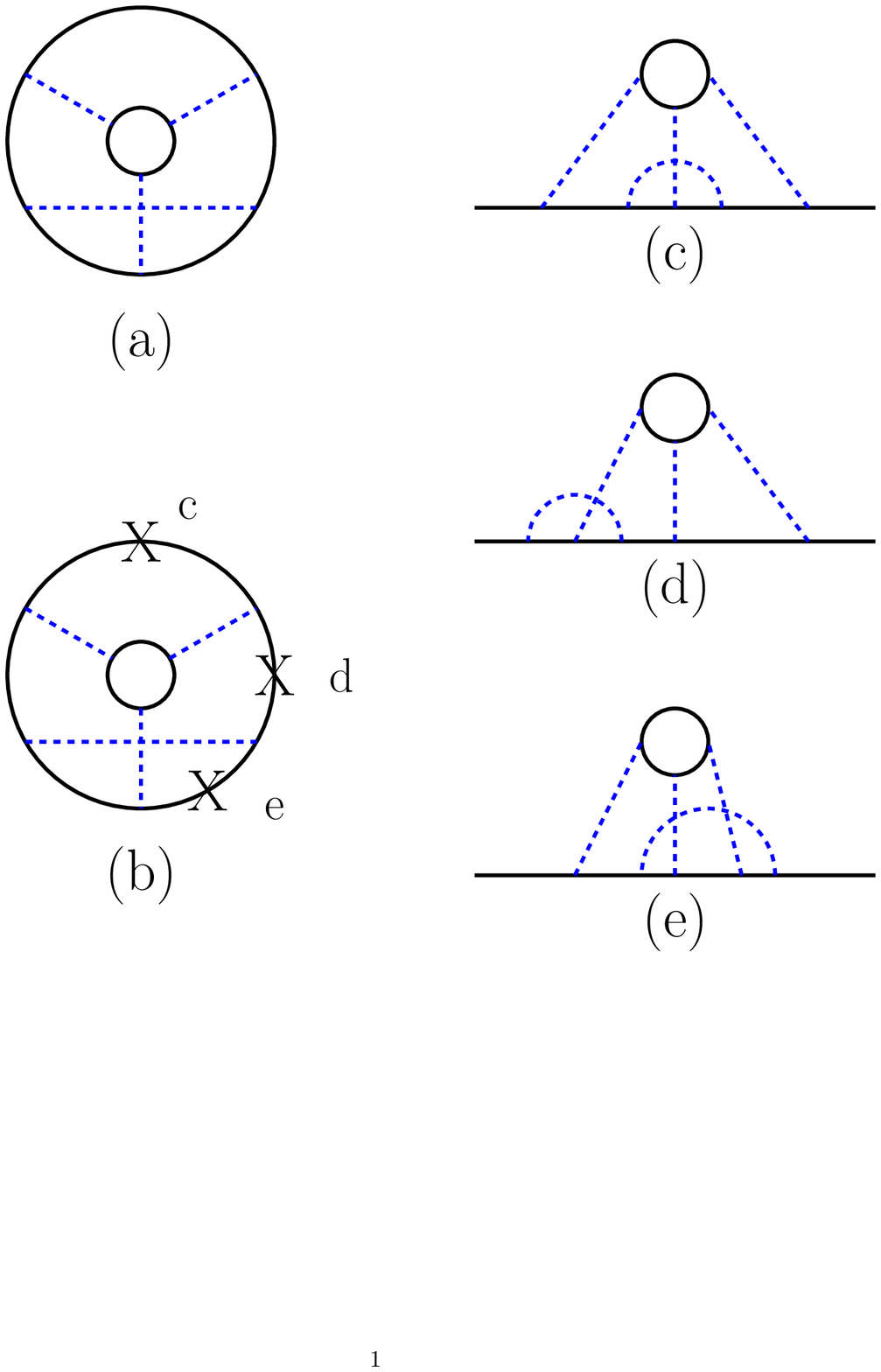}\vspace{1cm}  %
	\caption{An example of the high-order corrections.
(a) is the progenitor diagram, the possible contributions to the electric current appear when in (b) one of the crosses is substituted by the derivative $\partial_p Q_{0,W}$.
Using each of those crosses we form the diagram, which contributes to electric current.
Diagrams (d), (e), and (f) are the corresponding self-energy diagrams.
 }  %
	\label{fig_high-order}   %
\end{figure}

The higher order corrections may be considered in the similar way. The example of the higher order diagram is given in Fig.\ref{fig_high-order}. The sum of the Feynman diagrams represented in Fig.\ref{fig_high-order} (c), (d), (e) contribute the Fermion self energy that enters an expression for the total current presented in Fig. \ref{Fig_tadpole} (a) (the diagrams (d) and (e) are to be counted twice). The resulting contribution to the current is equal to the integral over momentum of the derivative of the progenitor diagram represented in Fig.\ref{fig_high-order} (a). This integral is zero for the system with  compact momentum space (say, when lattice regularization is used). The diagrams of Fig.\ref{fig_high-order} (c), (d), (e) appear when the diagram of  Fig.\ref{fig_high-order} (b) is cut at the positions of the crosses.

The obtained results mean the following:
(1) The interaction corrections to the current vanish.
(2) There is the following representation for the averaged  current along the $x$-axis in the considered system:
\begin {equation}\label{current_final}
{I}(g) = \int \frac{d^3 R d^3 p}{\beta S (2\pi)^3} Tr G_{g,W}(R,p) \star  \frac{\partial}{\partial p_x} Q_{g,W}(R,p)
\end{equation}
These two statements constitute the weak version of the Bloch theorem valid for the field - theoretical systems with gapped fermions. Namely, the persistent current in those systems being given by Eq. (\ref{current_final}) is the topological invariant, i.e. it is not changed when the system is modified smoothly (\rev{for the proof see Eq. (11) in \cite{Zhang+Zubkov2019_Bloch}}). All such systems may be divided into the homotopic classes. Within each class all systems are connected to each other by continuous deformation. If one system from the given homotopic class does not have a persistent current, then the other systems from the same class also cannot possess such a current. Interactions do not alter this conclusion according to the proof sketched above. In this prove the diagram technique proposed in the previous sections was used that operates with the Wigner transformed Green functions.

Notice that the proof does not rely on the precise expression for the Green functions.
We only used the property that the Fourier-transformed bosonic Green function ${\cal D}(q)$ is an even function of momentum. Therefore, the generalization of  our result to the case of the other interactions is straightforward.

\subsection{Quantum Hall conductivity}

In this subsection, we extend the above consideration to the calculation of radiative corrections to Hall conductivity. To some extent the results obtained below repeat those of  our previous work \cite{Zhang_2019_JETPL}.
We will consider the quantum Hall effect in the considered above system in the presence of magnetic field, electric field, and the interactions between the fermions due to exchange by scalar bosons.

The given two - dimensional system is subject to the periodic boundary condition, i.e. it is defined on the torus
shown in Fig.\ref{fig_torus}(b). Constant magnetic field is orthogonal to the plane. The
whole system is divided into the two parts: in region (I), i.e. $y\in [0,L]$, there is a constant electric field $E$ along the positive y direction, and there is an interaction between the fermions;
in region (II), i.e. $y\in (-L,0)$, there is a constant electric field $-E$
along the negative y direction, and there is no interaction between the fermions.
The lagrangian is given by
\begin {eqnarray}\label{Yukawa_a}
\mathcal{L}' =\bar{\psi} (( i \partial_{\mu}-A_{\mu})\gamma^{\mu} - m) \psi
                  + \partial_{\mu}\phi  \partial^{\mu} \phi - m^2_{\phi}\phi^2
                  -g \theta(y)\bar{\psi} \psi \phi,
\end{eqnarray}
where function $\theta(y)$ provides that the interaction exists only in the
region (I). Gauge field potential may be considered as the sum of two contributions  $A_{\mu}=A^{(m)}_{\mu}+A^{(e)}_{\mu}$ responsible for magnetic and electric fields correspondingly.
Eqs. (\ref{current}) to (\ref{deltaI}) still hold here,
while the self-energy function (in the leading order) is given by
\begin {eqnarray}\label{self-energy}
\Sigma_W (R,p) = \int \frac{d^3 q}{(2\pi)^3} G_{W}(R,p-q)D_W(R,q)
\end{eqnarray}
in which $D_W(R,q)$ (where $R = (\tau, x,y)$) is the Wigner transform of
$\theta(y_1)\theta(y_2)D(R_1-R_2)$, because the interaction only affects the region (I).
Despite of this difference, the approach of the last subsection still
works here, and the mentioned above weak version of the Bloch theorem holds here as well.
Namely, the total current \rev{averaged over the whole system} that consists of the two parts
(with and without interactions) does not depend on the value of $E$ (if it is sufficiently small). We have for the $x$ - component of electric current:
\rev{\begin{eqnarray}\label{current_total}
I_{tot}=(I_1+I_2)/2=(\bar{\sigma}_1 E+\bar{\sigma}_2 (-E))/2 + I_{tot}\Big|_{E=0},
\end{eqnarray}
where $I_1$ and $I_2$ are the average currents in the regions (I) and (II), respectively,
and $\bar{\sigma}_i$ is the corresponding average Hall conductivity. }
If, in addition, the system belongs to the homotopic class of the systems with vanishing equilibrium persistent current, then $I_{tot} =I_{tot}\Big|_{E=0} = 0$. However, for us it is sufficient, that $I_{tot}$ does not depend on $E$.
Now let us turn to the Hall conductivity.
Current density along the x-axis is given by
\begin {eqnarray}\label{current_density}
J(R) =\int \frac{d^3 p}{(2\pi)^3}  Tr G^{}_{W}(R,p) \frac{\partial}{\partial p_x} Q^{}_{W}(R,p).
\end{eqnarray}
In order to find out the Hall conductivity,
we assume, that electric part $A^{(e)}$ of electromagnetic potential is small (corresponding to external electric field), and  denote $A^{(e)} = \delta A$.
\rev{Therefore vector potential is $A^{(m)}_{\mu}+ \delta A_{\mu}$. }
In the following  we omit subscript $(m)$ for brevity. We consider variation of electric current $\delta J$ with respect to the variation $\delta A$, and obtain
\begin{eqnarray}\label{current_a}
J[A+\delta A] = \int \frac{d^3 p}{(2\pi)^3}  Tr G_{W}(R,p)  \frac{\partial}{\partial p_x} {\cal Q}(p-{ A}(R)-\delta { A}),
\end{eqnarray}
where $G_{W}$ satisfies  $G_{W}(R,p) \star Q_{W}(R,p)=1$, with
$Q_{W}(R,p)= {\cal Q}(p-{ A}(R)-\delta { A})$ \cite{Suleymanov_2019}.
Notice that Eq.(\ref{current_a}) does not contain the $ \star$ operation.
Using the expansion in powers of $\delta A$ we obtain ${\cal Q}(p-{A}(R)-\delta { A})\approx
{\cal Q}(p-{A}(R))-\partial^{\mu}{\cal Q}\delta {A}_{\mu}$, next we
 expand function $G_{W}(R,p)$ in powers of $\delta A$:   $G_{W}(R,p)=G^{(0)}_{W}+G^{(1)}_{W}+...$,
with $G^{(n)}_{W}\sim O([\delta A]^n)$.
In the leading (zeroth) order  $G^{(0)}_{W}$ satisfies equation
\begin {eqnarray}\label{WignerEqu_0}
G^{(0)}_{W}(R,p)  \star {\cal Q}(p-{ A}(R))= 1.
\end{eqnarray}
%
In the next (the first) order
$G^{(1)}_{W}$ satisfies
\begin {eqnarray}\label{WignerEqu_1}
0=G^{(1)}_{W}(R,p) \star {\cal Q}(p-{A}(R)) -
    G^{(0)}_{W}(R,p) \star \Big(\frac{\partial {\cal Q}(R,p)}{\partial p_{\mu}} \delta {A}^{\mu}\Big)
\end{eqnarray}
Solution of this equation gives
\begin {eqnarray}\label{WignerEqu_1}
G^{(1)}_{W}(R,p)&=&G^{(0)}_{W}(R,p) \star
            \Big(\frac{\partial {\cal Q}(R,p)}{\partial p_{\mu}} \delta {\cal A}^{\mu}\Big)
             \star G^{(0)}_{W}(R,p)    \nonumber\\
                        &=&(G^{(0)}_{W} \star \partial_{\mu} {\cal Q} \star G^{(0)}_{W}) \delta{\cal A}^{\mu}
                          +\frac{i}{2}(\partial_{\mu} G^{(0)}_{W} \star \partial_{\nu} {\cal Q} \star G^{(0)}_{W}) \delta F_{\mu\nu}
\end{eqnarray}
Therefore, we find the variation of current $\delta J $, up to the linear term in  $\delta A$:
\begin {eqnarray}\label{current_b}
\delta J  =\frac{i}{2}\delta F_{lm}  \int \frac{d^3 p}{(2\pi)^3}Tr[ \partial_l G^{(0)}_{W}
            \star \partial_m Q^{(0)}_{W} \star G^{(0)}_{W})] \partial_{p_x} Q^{(0)}_{W}.
\end{eqnarray}
%
Then we obtain  expression for  local Hall conductivity:
\begin {eqnarray}\label{conductivity}
\sigma  =\frac{1}{2}  \int \frac{d^3 p}{(2\pi)^3}Tr[ \partial_{[p_y} G^{(0)}_{W}
           \star  \partial_{\omega]} Q^{(0)}_{W}\star G^{(0)}_{W})] \partial_{p_x} Q^{(0)}_{W}.
\end{eqnarray}
%
\rev{We assume that the electric field strength expressed by $\delta F_{lm}$ is constant within each region ((I) and (II)),
but has different signs in those regions. Therefore, the total current is
\begin {eqnarray}\label{conductance}
I_{tot} =\int \delta J \frac{dxdy}{2{\cal S}}
          =  E\int_{R\in ({\rm I})} \frac{dxdy}{2{\cal S}}  \sigma_1 (g)
            - E \int_{R\in ({\rm II})} \frac{dxdy}{2{\cal S}} \sigma_2 (0) +I_{tot}\Big|_{E=0}
\end{eqnarray}
Since $I_{tot}$ does not depend on $E$ (it is the topological invariant, see the previous subsection), we know that the Hall conductivities (averaged over each region) satisfy
$\bar{\sigma}_1(g)=\bar{\sigma}_2(0)$.
Futhermore, from $\bar{\sigma}_1(0)=\bar{\sigma}_2(0)$, one can find out that $\bar{\sigma}_1(0)=\bar{\sigma}_1(g)$.}
Now, we come to the expression for the Hall conductivity averaged over the area of the part $(I)$, where the interactions are present:
\begin{eqnarray}
\bar{\sigma}_1&=&\frac{1}{2}\int_{R\in ({\rm I})} \frac{dxdy}{{\cal S}} \int \frac{d^3 p}{(2\pi)^3}Tr[ \partial_{[p_y} G^{(0)}_{W}
           \star  \partial_{\omega ]} Q^{(0)}_{W}\star G^{(0)}_{W})] \partial_{p_x} Q^{(0)}_{W}\\
&=&\frac{1}{2}\int_{R\in ({\rm I})}\frac{dxdy}{{\cal S}}   \int \frac{d^3 p}{(2\pi)^3}Tr[ \partial_{[p_y} G^{(0)}_{W}
           \star  \partial_{\omega]} Q^{(0)}_{W}\star  G^{(0)}_{W})]\star  \partial_{p_x} Q^{(0)}_{W}
\end{eqnarray}
It does not depend on the value of $g$.
In the second line of the above equation, we change ordinary product into the $\star$  product, because the terms standing under the integral do not depend on electric field. Therefore, the periodic boundary conditions may be imposed.

The last expression allows us to obtain the following representation for the average Hall  conductivity (Electric field is directed along the $y$ axis):
$\bar{\sigma}_{xy} = {\cal N}/ 2 \pi$,
where ${\cal N}$ is the topological invariant in phase space, which is the generalization of the  classical TKNN invariant \cite{TKNN}
\begin{eqnarray}
 {\cal N} =   \frac{T}{{ S}\,3!\,4\pi^2}\,  \epsilon_{ijk} \,
\int  d^3 x  \, \int d^3p \, Tr
  {G}_W(p,x)\star  \frac{\partial {Q}_W(p,x)}{\partial p_i} \star
 \frac{\partial  {G}_W( p,x)}{\partial p_j} \star  \frac{\partial  {Q}_W(p,x)}{\partial p_k}
	\label{calM2d23c}
\end{eqnarray}
In this expression $\delta A$ is set equal to zero. According to the above results Eq. (\ref{calM2d23c}) does not depend on $g$ to all orders of the perturbation theory in $g$. The latter result is obtained using the particular case of the diagram technique developed in the previous sections.


\section{Conclusions}

In the present paper we constructed the diagram technique for the non - homogeneous model with Dirac fermion interacting with the scalar field. Both types of excitations propagate in the presence of the inhomogeneous background provided by the external gauge field. We consider arbitrary Green functions in this model corresponding to the external fermion legs. The obtained construction may easily be generalized to the case of the other models existing in the inhomogeneous background. The essential feature of our technique is that all diagrams are expressed through the Wigner transformed propagators. Those depend on both coordinates $R$ and momenta $p$. The main advantage of the proposed construction is that the momenta in the diagrams are "conserved", i.e. in each vertex the same rules are valid as in the homogeneous theory: the sum of the incoming and outgoing momenta is equal to zero. Each propagator carries one momentum. This reveals the one - to one correspondence with the corresponding homogeneous model. The inhomogeneity of the theory is encoded in the dependence of the propagators on $R$, and in the replacement of ordinary products by the Moyal products of Wigner - Weyl calculus. This construction has been used in our two previous papers \cite{Zhang_2019_JETPL} and \cite{ZZ2019_3}. In those papers the particular case of the diagram technique of the present paper has been developed for the  particular cases  of the diagrams. It allows to prove the non - renormalization of Hall conductivity by weak interactions, and the non - renormalization of total persistent current by weak interactions for the model of massive fermions.  Here we generalize the constructions of \cite{Zhang_2019_JETPL} and \cite{ZZ2019_3} to arbitrary forms of the diagrams. \rev{We also repeat briefly in the present paper the considerations of  \cite{Zhang_2019_JETPL} and \cite{ZZ2019_3} adopted to the model with the action of Eq. (\ref{Yukawa}) as an application of the proposed diagram technique.}

The other possible applications of our technique may exist to the physics of high-energy heavy-ion collisions, when the two atomic nuclei collide at relativistic energies and
 generate strong electromagnetic fields \cite{Deng+Huang_2016}.
If the collision energy surpasses a certain threshold,
the collision produces quark-gluon plasma (see \cite{Zubkov_2018} and references therein). The behavior of the
charged particles is affected by those strong electromagnetic fields. One can say, that various excitations move in the inhomogeneous background given by the mentioned electromagnetic fields. Various non-dissipative transport effects are expected to be observed in the heavy-ion collisions. Those are, for example, the chiral magnetic effect, the chiral separation effect, the chiral vortical effect. In the presence of the homogeneous magnetic field/homogeneous rotation angular velocity the corresponding conductivities are given by the topological invariants in momentum space (see \cite{Zubkov_2018,KZ2018_2,ZK2018,AKZ2018,Z2018,ZK2017,KZ2017,Z2016_QCD}). It is expected that in the presence of the inhomogeneous background (given by the inhomogeneous magnetic field, inhomogeneous rotation, etc) the corresponding conductivities will be given by the topological invariants in phase space composed of the Wigner transformed Green functions. This expectation is based on the recent extension of the topological representation for the Hall conductivity to the case of inhomogeneous magnetic field \cite{Zhang_2019_JETPL}. The representation of the non - dissipative conductivities in terms of the Wigner transformed propagators may repeat the corresponding homogeneous constructions with the ordinary product replaced by the Moyal product. The topological nature of the quantities simplifies the use of the Moyal product and may allow us to consider relatively easily the interaction corrections to those effects (as in \cite{Zhang_2019_JETPL} and \cite{ZZ2019_3}). Possible applications of the proposed technique may be found also in various problems of condensed matter physics, where the inhomogeneity may be caused by many factors including the elastic deformations. Here as well the consideration of various non - dissipative transport phenomena (fractional quantum Hall effect, spin Hall effect, etc) may benefit from the use of our diagram technique.

The authors are grateful for useful discussions to I.Fialkovsky, M.Suleymanov, and Xi Wu.




\end{document}